\documentclass[%
 reprint,
superscriptaddress,
nofootinbib,
 amsmath,amssymb,
 aps,
]{revtex4-1}

\usepackage{aas_macros}
\usepackage{graphicx}
\usepackage{subcaption}

\newcommand{\jstat}{$\mathcal{J}$-statistic}
\newcommand{\fstat}{$\mathcal{F}$-statistic}
\newcommand{\cstat}{$\mathcal{C}$-statistic}
\newcommand{\Tsft}{T_\mathrm{SFT}}
\newcommand{\Tobs}{T_\mathrm{obs}}
\newcommand{\Tdrift}{T_\mathrm{drift}}
\newcommand{\fstar}{f_\star}
\newcommand{\fdrift}{\Delta f_{\mathrm{drift}}}
\newcommand{\fband}{\Delta f_{\mathrm{band}}}
\newcommand{\fline}{f_{\mathrm{line}}}

\newcommand{\Sth}{S_\mathrm{th}}
\newcommand{\Tasc}{T_{\mathrm{asc}}}
\newcommand{\heff}{h_0^{\mathrm{eff}}}
\newcommand{\heq}{h_0^{\mathrm{eq}}}
\newcommand{\hnfive}{h_0^{95\%}}

\begin{document}


\title{Search for gravitational waves from Scorpius X-1 in the second Advanced LIGO observing run with an improved hidden Markov model}

\author{B.~P.~Abbott}
\affiliation{LIGO, California Institute of Technology, Pasadena, CA 91125, USA}
\author{R.~Abbott}
\affiliation{LIGO, California Institute of Technology, Pasadena, CA 91125, USA}
\author{T.~D.~Abbott}
\affiliation{Louisiana State University, Baton Rouge, LA 70803, USA}
\author{S.~Abraham}
\affiliation{Inter-University Centre for Astronomy and Astrophysics, Pune 411007, India}
\author{F.~Acernese}
\affiliation{Universit\`a di Salerno, Fisciano, I-84084 Salerno, Italy}
\affiliation{INFN, Sezione di Napoli, Complesso Universitario di Monte S.Angelo, I-80126 Napoli, Italy}
\author{K.~Ackley}
\affiliation{OzGrav, School of Physics \& Astronomy, Monash University, Clayton 3800, Victoria, Australia}
\author{C.~Adams}
\affiliation{LIGO Livingston Observatory, Livingston, LA 70754, USA}
\author{R.~X.~Adhikari}
\affiliation{LIGO, California Institute of Technology, Pasadena, CA 91125, USA}
\author{V.~B.~Adya}
\affiliation{Max Planck Institute for Gravitational Physics (Albert Einstein Institute), D-30167 Hannover, Germany}
\affiliation{Leibniz Universit\"at Hannover, D-30167 Hannover, Germany}
\author{C.~Affeldt}
\affiliation{Max Planck Institute for Gravitational Physics (Albert Einstein Institute), D-30167 Hannover, Germany}
\affiliation{Leibniz Universit\"at Hannover, D-30167 Hannover, Germany}
% ...

\collaboration{The LIGO Scientific Collaboration and the Virgo Collaboration}

\author{L.~M.~Dunn}
\affiliation{OzGrav, University of Melbourne, Parkville, Victoria 3010, Australia}

\author{S.~Suvorova}
\affiliation{OzGrav, University of Melbourne, Parkville, Victoria 3010, Australia}

\author{R.~J.~Evans}
\affiliation{OzGrav, University of Melbourne, Parkville, Victoria 3010, Australia}

\author{W.~Moran}
\affiliation{OzGrav, University of Melbourne, Parkville, Victoria 3010, Australia}

\begin{abstract}
	We present results from a semicoherent search for continuous gravitational
	waves from
	the low-mass X-ray binary Scorpius X-1, using a hidden Markov model (HMM) to
	track spin wandering.
	This search improves on previous HMM-based searches of LIGO data by using
	an improved frequency domain matched filter, the \jstat{}, and by
	analysing data from Advanced LIGO's second observing run.
	In the frequency range searched, from $60$ to $650\,\mathrm{Hz}$,
	we find no evidence of gravitational
	radiation. At $194.6\,\mathrm{Hz}$, the most sensitive search frequency, we report
	an upper limit on gravitational wave strain (at 95\% confidence) of $h_0^{95\%}
	= 3.47 \times 10^{-25}$ when marginalising over source inclination angle.
	This is
	the most sensitive search for Scorpius X-1, to date, that is specifically
	designed to be robust
	in the presence of spin wandering.
\end{abstract}

\maketitle

\section{Introduction}
Rotating
neutron stars
with non-axisymmetric deformations
are predicted to emit persistent, periodic gravitational
radiation. They are a key target for continuous-wave
searches performed with gravitational wave (GW) detectors such as the
second-generation Advanced Laser Interferometer Gravitational-wave Observatory
(Advanced LIGO)
\citep{2013riles,2010harry,2015lscadvligo,2015acernese,2011andersson} and Virgo
\citep{2015acernese}.
The time-varying quadrupole moment necessary for GW emission may result from
thermal \citep{2000ushomirsky,2013johnsonmcdaniel},
or magnetic \citep{2002cutler,2011mastrano,2013lasky} gradients,
$r$-modes \citep{2002heyl,2003arras,2009bondarescu},
or nonaxisymmetric
circulation of the superfluid interior \citep{2006peralta,2008vaneysden,2010bennett,2015melatosdouglass}.
These mechanisms produce
signals at certain multiples of the spin frequency $\fstar$ \citep{2013riles}.
Of particular interest are
accreting low-mass X-ray binaries (LMXB),
such as Scorpius X-1 (Sco~X-1), where a neutron star is spun up by accretion
from its stellar companion.
Electromagnetic observations of LMXBs
to date imply $f_\star \lesssim 620\,\mathrm{Hz}$ \citep{2003chakrabarty},
well short of the theoretical
centrifugal break-up limit $f_\star \lesssim 1.5\,\mathrm{kHz}$ \citep{1994cook}.
Regardless of the
exact GW mechanism, the latter observation suggests an
equilibrium between the spin-up accretion torque, and
GW spin-down torque
\citep{1998bildsten,1978papaloizoupringle,1984wagoner}.
Torque balance also implies a
relation between X-ray luminosity and the GW strain,
making Sco~X-1, the brightest LMXB X-ray source, the most promising known target.

Initial LIGO,
a first-generation detector,
started taking science data in 2002.
It reached its design sensitivity
in Science Run
5 (S5), starting 2005 \citep{2009lscinitialligo},
and exceeded it in
Science Run 6 (S6) \citep{2012lscS6}. Following detector
upgrades, the second-generation Advanced LIGO interferometer \citep{2010harry} began taking science
data during Observing Run 1 (O1), which ran from September 2015 to January
2016.
The strain noise in O1 is three to four times lower than S6 between
$100\,\mathrm{Hz}$ and $300\,\mathrm{Hz}$ \citep{2016lscO1}. During this
period, LIGO observed
three binary black hole mergers, GW150914 \citep{2016lscgw150914}, GW151012
and GW151226 \citep{2016lscgw151226}.
Observing Run 2 (O2)
began on November 2016, and ran until 26 August 2017. From 1 August 2017, the
two LIGO detectors were joined by Virgo, resulting in a three-detector network.
As well as further binary
black hole mergers \citep{2018o2catalogue}, LIGO and Virgo made the first gravitational wave
observation of a binary neutron-star
merger during O2 \citep{2017lsc170817}.

No search has yet reported a detection of a continuous wave source.
To date, four searches for Sco~X-1 have been conducted on Initial LIGO data,
and three on Advanced LIGO data.
The first search coherently analysed the
most-sensitive six hour segment from Science Run 2 (S2) using the \fstat{} \citep{1998jks}, a
maximum likelihood detection statistic \citep{2007lscfstat}. The second was a directed,
semi-coherent analysis using the \cstat{} \citep{2015lsccstat}.
The third, also a directed analysis, used
the TwoSpect algorithm
on doubly Fourier transformed S5 data
\citep{2011goetz,2014lsctwospect,2016meadors}.
The fourth applied the radiometer algorithm \citep{2006ballmer} to conduct a directed
search on S4 \citep{2007lscradiometerS4},
S5 \citep{2011lscradiometerS5},
and later O1 \citep{2017lscradiometerO1} data.
Three LMXB searches have been performed with Advanced LIGO data, comprising the
radiometer search \citep{2017lscradiometerO1}, an analysis based on a hidden Markov model (HMM)
\citep{2017lscHMM}, and a cross-correlation analysis
\citep{2015whelan,2008dhurandhar,2017lscO1crosscorr}.
The upper limits established by these searches are summarized in Table~\ref{tbl:previoussrch}.

Astrophysical modeling and X-ray observations suggest
that the spin frequency of an LMXB
wanders stochastically in response to fluctuations in the
hydromagnetic accretion torque
\citep{1993dekool,1993baykal,1997bildsten,2008watts}. As no
electromagnetic measurements of $f_\star$ are available
to guide a gravitational wave search
for Sco~X-1, such searches
must either account for spin wandering or limit their observing times and/or
coherence times in accordance with the anticipated timescale and amplitude of
the spin wandering \citep{2018mukherjee}.
For example, the sideband search
described in Ref.~\citep{2015lsccstat} is restricted to data segments no longer than ten
days. The hidden Markov model (HMM) tracker,
first applied to the search for Sco~X-1 in Ref.~\citep{2017lscHMM}, is an effective
technique for detecting the most probable underlying
spin frequency, $f_\star(t)$ and thus accounting for spin wandering.

The signal from an binary source is
Doppler shifted, as the neutron star revolves around the barycentre of the
binary, dispersing power into orbital sidebands 
near
the source frame emission frequency. The separation of
these sidebands and the source-frame frequency depends on the binary orbital
parameters and $f$, but is typically within
0.05 per cent of the gravitational wave frequency
for a source such as Sco~X-1.
Four maximum-likelihood matched filters have been developed to detect these
sidebands: the \cstat{}, which weights sidebands equally \citep{2015lsccstat},
the binary modulated \fstat{} \citep{2015leaci},
the Bessel-weighted \fstat{} \citep{2016suvorova}, and the
\jstat{}, which extends the
Bessel-weighted \fstat{} to account for the phase of the binary orbit
\citep{2017suvorova}.
Any of these matched filters can be combined with the HMM to conduct a search
for signals from a binary source that accounts for spin wandering.

In this paper, we combine the \jstat{} described in Ref.~\citep{2017suvorova} with the HMM
described in that paper and Refs.~\citep{2016suvorova}
and~\citep{2017lscHMM}, and perform a directed search of Advanced LIGO O2 data
for evidence of a gravitational wave signal from Sco~X-1.
In the search band
60--650 Hz, we find no evidence of a gravitational wave signal.
The paper is organized as follows. In Section~II, we briefly review the HMM and
the \jstat{}. In Section~III, we discuss the search strategy and parameter
space.  In Section~IV, we report on the
results from the search and veto candidates corresponding to instrumental
artifacts.
In Section~V, we discuss the search sensitivity and
consequent upper limits
on the gravitational wave strain.

\begin{table*}
	\caption{Summary of indicative upper limits achieved in previous searches for Sco~X-1. VSR2 and VSR3 are Virgo Science Runs 2 and 3, respectively.  Where applicable, the upper limits refer to signals of unknown polarization.}
	\label{tbl:previoussrch}
	\begin{tabular}{llll}
		Search            & Data & Upper limit & Reference \\
		\hline
		         \fstat{} & S2   & $h_0^{95\%} \lesssim 2\times10^{-22}$ at 464--484 Hz, 604 -- 626 Hz & \citep{2007lscfstat} \\
		\cstat{}          & S5   & $h_0^{95\%} \lesssim 8 \times 10^{-25}$ at 150 Hz & \citep{2015lsccstat} \\
		TwoSpect          & S6, VSR2, VSR3 & $h_0^{95\%} \lesssim 2 \times 10^{-23}$ at 20 -- 57.25 Hz & \citep{2014lsctwospect} \\
		Radiometer        & S4, S5 & $h_0^{90\%} \lesssim 2 \times 10^{-24}$ at 150 Hz & \citep{2011lscradiometerS5,messengerRadiometerConversion} \\
		TwoSpect          & S6     & $h_0^{95\%} \lesssim 1.8 \times 10^{-24}$ at 165 Hz & \citep{2017meadors} \\
		Radiometer        & O1     & $h_0^{90\%} \lesssim 6.7 \times 10^{-25}$ at 130 -- 175 Hz & \citep{2017lscradiometerO1} \\
		Viterbi 1.0       & O1     & $h_0^{95\%} \lesssim 8.3 \times 10^{-25}$ at 106 Hz & \citep{2017lscHMM} \\
		Cross correlation & O1     & $h_0^{95\%} \lesssim 2.3 \times 10^{-25}$ at 175 Hz & \citep{2017lscO1crosscorr} \\
		\hline
	\end{tabular}
\end{table*}

\section{Search algorithm}
In this section, we 
outline the two key components of the search algorithm:
the HMM, used to recover the most
probable
spin history $f_0(t)$,
and the \jstat{}, the matched filter that
accounts for
the Doppler shifts introduced by the orbital motions of the Earth and the
LMXB.
The HMM formalism is the same as
used in Refs.~\citep{2017lscHMM,2016suvorova,2017suvorova}, so we review it
only briefly. The \jstat{} is described fully in Ref.~\citep{2017suvorova};
again, we review it briefly.

\subsection{HMM formalism}
\label{ssec:hmmformalism}
A Markov model
describes a stochastic process in terms of
a state variable $q(t)$, which
transitions between allowable states $\{ q_1, \cdots, q_{N_Q} \}$ at
discrete times $\{ t_0, \cdots, t_{N_T} \}$.
The transition matrix $A_{q_jq_i}$
represents the probability of jumping from state $q_i$ at the time $t = t_n$ to
$q_j$ at $t = t_{n + 1}$ depending only on $q(t_n)$.
A HMM extends the Markov model
to situations where
direct observation
of $q(t)$ is impossible
[$q(t)$ is called the hidden state].
Instead one measures
an observable state $o(t)$,
selected from $\{ o_1, \cdots, o_{N_o} \}$, which is
related to the hidden state by the
emission matrix $L_{o_jq_i}$, which gives the likelihood that the system is in
state $q_i$ given the observation $o_j$.
In gravitational wave searches for LMXBs like Sco~X-1, where the spin frequency
cannot be measured electromagnetically, it is natural to map $q(t)$ to
$f_0(t)$ and $o(t)$ to the raw interferometer data, some equivalent
intermediate data product (e.g. short Fourier transforms), or a detection
statistic (e.g., \fstat{}, \jstat{}).

In an LMXB search,
we divide the total observation (duration $\Tobs$)
into $N_T$ equal segments of length $\Tdrift = \Tobs / N_T$.
In practice, $\Tdrift$ is chosen on astrophysical grounds to give $N_T = \lceil
\Tobs / \Tdrift
\rceil$
based on an estimates of plausible spin-wandering timescales
\citep{2018mukherjee}; in this paper we follow Ref.~\citep{2017lscHMM} in
choosing $\Tdrift = 10\,\mathrm{d}$.
The tracker is able to track the signal even if the spin frequency occasionally
jumps by two bins as it can catch up to the signal path,
although with an attendant loss of
sensitivity as the recovered must include a step that contains only noise.

In each segment, the emission probability $L_{o_jq_i}$ is computed from some
frequency domain estimator $G(f)$ such as the maximum likelihood $\mathcal{F}$- or
\jstat{} (discussed in
Section~\ref{ssec:jstat}). The frequency resolution of the estimator
is $\fdrift =
1/(2\Tdrift)$.
The probability that an observation $O = \{o(t_1), \cdots, o(t_{N_T})\}$ is
associated with a particular hidden path $Q = \{ q(t_0), \cdots, q(t_{N_T}) \}$
is then given by:
\begin{align}
	\notag
	P(Q|O) \propto &L_{o(t_{N_T})q(t_{N_T})} A_{q(t_{N_T})q(t_{N_T-1})} \cdots
	L_{o(t_1)q(t_1)}\\
	&\times A_{q(t_1)q(t_0)} \Pi_{q(t_0)},
\end{align}
where $\Pi_{q_i}$ is the prior, i.e., the probability the system starts in state
$q_i$ at $t = t_0$. For this search, we take a flat prior. (Note that there is
no initial observation $o(t_0)$ as the initial state of the system is captured
by the prior.)
The task, then, is to find the
optimal hidden path $Q^\star$, that is, the path $Q^\star$ that maximizes
$P(Q|O)$ given $O$.
We find $Q^\star$ efficiently with the recursive Viterbi
algorithm \citep{1967viterbi},
which is discussed in detail in Appendix~A of Ref.~\citep{2017lscHMM}.

In this paper, we follow the convention in Ref.~\citep{2017lscHMM} of defining
the Viterbi detection score $S$ for a path as the number of standard
deviations by which that path's log likelihood exceeds the mean log likelihood
of all paths. Mathematically we have
\begin{align}
	S = \frac{\ln \delta_{q^\star}(t_{N_T}) - \mu_{\ln\delta(t_{N_T})}}
	  {\sigma_{\ln\delta(t_{N_T})}},
	  \label{eqn:viterbiscore}
\end{align}
where
\begin{align}
	\mu_{\ln\delta(t_{N_T})} = N_Q^{-1} \sum_{i=1}^{N_Q}
	  \ln \delta_{q_i} (t_{N_T}),
\end{align}
\begin{align}
	\sigma^2_{\ln \delta(t_{N_T})} = N_Q^{-1} \sum_{i=1}^{N_Q}
	 [\ln \delta_{q_i(t_{N_T})} - \mu_{\ln\delta(t_{N_T})}]^2,
\end{align}
$\delta_{q_i}(t_{N_T})$ denotes the likelihood of the most likely path
ending in state $q_i$ at step $N_T$, and $\delta_{q^\star}(t_{N_T}) =
\max_i\delta_{q_i}(t_{N_T})$ is the
likelihood of the optimal path overall.

\subsection{\jstat}
\label{ssec:jstat}
The frequency domain estimator $G(f)$ converts the
interferometer data into the
likelihood that a signal is present at frequency $f$.
For a continuous-wave search for an isolated neutron star, the
maximum-likelihood \fstat{} \citep{1998jks} is a typical choice for $G(f)$. The \fstat{} accounts
for the diurnal rotation of the Earth, and its orbit around
the Solar System barycentre. It is an almost optimal matched filter for a
biaxial rotor \citep{2009prix}.

For a neutron star in a binary system, such as an LMXB, the
signal is frequency (Doppler) modulated by the binary orbital motion as well.
Ref.~\citep{2017lscHMM} used the Bessel-weighted \fstat{} to
account for this modulation, without using information about the orbital phase.
Ref.~\citep{2017suvorova} introduced the
\jstat{}, which is a matched filter that extends the \fstat{} to
include orbital phase in the signal model.
The orbital Doppler effect
distributes the \fstat{} power into approximately
$2m + 1$ orbital sidebands separated by $P^{-1}$,
with $m = \left\lceil 2\pi f_\star a_0 \right\rceil$, where
$\left\lceil \cdot \right\rceil$ denotes rounding up to the nearest integer,
$P$ is the orbital period and $a_0 = (a\sin i)/c$
is the light travel time across the projected semi-major axis
(where $a$ is semi-major axis
and $i$ is the inclination angle of
the binary).
For
a zero-eccentricity Keplerian orbit, the
Jacobi-Anger identity may be used to
expand the signal
$h(t)$ in terms of Bessel
functions, suggesting a matched filter of the form \citep{2016suvorova,2017suvorova}
\begin{align}
	G(f) = \mathcal{F}(f) \otimes B(f),
\end{align}
with
\begin{align}
	\label{eqn:convolution}
	B(f) = \sum_{s = -m}^m J_s(2\pi f_0 a_0) e^{-is\phi_a} \delta(f - s/P),
\end{align}
where $J_s(z)$ is the Bessel function of the first kind of order $s$,
$\phi_a$ is the orbital phase at a reference time, and
$\delta$ is the Dirac delta function.

All else being equal, using the \jstat{} instead of the Bessel-weighted
\fstat{} improves sensitivity by a factor of approximately four.
Ref.~\citep{2017suvorova}, particularly Section~IV of that paper, examines the
difference between the two estimators in depth.

The Bessel-weighted \fstat{} requires a search
over $a_0$ but does not depend on $\phi_a$. By contrast, the more-sensitive
\jstat{} involves searching over $\phi_a$ too. In this paper we apply the
\jstat{} to search for Sco~X-1. Details of the search and priors
derived from
electromagnetic measurements are discussed in Section \ref{sec:search}.

\section{LIGO O2 search}
\label{sec:search}

\subsection{Sco~X-1 parameters}
\label{ssec:scox1params}
The matched filter described in Section~\ref{ssec:jstat} depends on three
binary
orbital parameters: the period $P$, the projected
semi-major axis $a_0$ and the phase $\phi_a$. The \fstat{} depends
on the sky location $\alpha$ (right ascension) and $\delta$ (declination), and
optionally the source frequency derivatives. For this search, we assume there
is no secular evolution in frequency.
The other parameters have
been measured electromagnetically for Sco~X-1 and are presented in
Table~\ref{tbl:scox1params}.

For $\alpha$, $\delta$ and $P$, the uncertainties in the electromagnetic
measurements are small enough that they have no appreciable effect on the
sensitivity of the search \citep{2014sammut,2015lscSidebandScoX1,2015leaci},
and a single, central value can be assumed.
However, the uncertainties in $a_0$ and $\phi_a$
cannot be neglected.
The time spent
searching orbital parameters scales as the number of ($a_0$, $\phi_a$) pairs.
Careful
selection of the ranges of $a_0$ and $\phi_a$ is essential to keep computational
costs low.

The previous analysis, described in Ref.~\citep{2017lscHMM}, used the
Bessel-weighted \fstat{} in place of the \jstat{}, and searched over a
uniformly-gridded range of $a_0$, where the grid resolution did not depend on
frequency.
However, the \jstat{} is more sensitive to mismatch in the binary orbital
parameters, so a finer grid is required.
We must also choose an appropriate grid for $\phi_a$.
(The Bessel-weighted \fstat{} is independent of $\phi_a$.)

As the \jstat{} has a similar overall response to parameter mismatches as the
binary \fstat{}, we follow the formalism in Ref.~\citep{2015leaci} to select
an appropriate parameter space gridding.
We choose a grid which limits the
maximum loss in signal-to-noise ratio (mismatch) $\mu_{\mathrm{max}}$ to
$\mu_{\mathrm{max}} = 0.1$.
Equation~(71) in Ref.~\citep{2015leaci} gives a general equation for the number
of grid points needed for each search parameter. For the particular search
considered in this paper, the number of choices for $a_0$ and $\phi_a$
are
\begin{align}
	N_{a_0} = \left\lceil \frac{\pi\sqrt{2}}{2}\mu^{-1/2}_{\mathrm{max}} f_0 \Delta a_0 \right\rceil,
\end{align}
\begin{align}
	N_{\phi_a} = \left\lceil
	\frac{1}{2}\mu^{-1/2}_{\mathrm{max}} f_0 a_0 \left( \frac{2\pi}{P} \right) \Delta \phi_a
	\right\rceil
	,
\end{align}
where $\Delta a_0$ and $\Delta \phi_a$ are the widths of the search ranges for $a_0$ and
$\phi_a$ respectively.
The number of orbital parameters to be searched depends on the search
frequency.
Accordingly for each search sub-band, we adopt a different grid resolution,
with the grid refined at higher frequencies.
In the sub-band beginning at
60 Hz, we have $N_{a_0} = 768$ and $N_{\phi_a} = 78$; in the sub-band beginning
at 650
Hz, we have $N_{a_0} = 8\,227$ and $N_{\phi_a} = 824$.
In principle we could achieve
further computational savings by noting that $N_{\phi_a}$ also depends on $a_0$,
but for safety
we use the largest $a_0$.

The search range for $a_0$ is $1.45 \leq a_0/(1\,\mathrm{s}) \leq
3.25$, which matches the most
recent electromagnetic measurement
\citep{WANG_NEW_SCOX1} and widens the error bars on the
widely-cited and previous best published measurement, $a_0 = 1.44 \pm
0.18\,\mathrm{s}$.
\citep{2002steeghs}.

The orbital phase $\phi_a$ can be related to the
elecromagnetically measured time of
ascension, $\Tasc$, given in Table~\ref{tbl:scox1params}, by
\begin{align}
	\phi_a = 2 \pi \Tasc/P \pmod{2 \pi}.
\end{align}
The one-sigma uncertainty in the published value for $\Tasc$ is $\pm 50\,\mathrm{s}$
\citep{WANG_NEW_SCOX1,2015messenger}, for a time of ascension at GPS time
$974\,416\,624\,\mathrm{s}$ (in November 2010).
As O2 took place significantly after
this time, to make a conservative estimate on appropriate error bars
for $\Tasc$,
we advance $\Tasc$ by adding $3\,135$ orbital
periods to the time of ascension taken from Ref.~\citep{WANG_NEW_SCOX1}.
As
there is uncertainty associated with the measured orbital period, this widens
the one-sigma uncertainty of $\Tasc$ to $\pm 144\,\mathrm{s}$, which
we round up to $\pm 150\,\mathrm{s}$.
To cover a significant portion of the measured $\Tasc$ range while keeping the
search computationally feasible,
we search a two-sigma range around the central
$\Tasc$, namely,
$1\,164\,543\,014 \leq \Tasc/(1 \,\mathrm{s}) \leq 1\,164\,543\,614$ (expressed
for presentation purposes
as the time of the last ascension before the start of O2).

As there is no electromagnetic measurement of $f_\star$ for Sco~X-1,
we search the band $60 \leq f_\star / (1\,\mathrm{Hz}) \leq  650$,
where LIGO is most sensitive, again adopting a uniform prior (see
Section~\ref{ssec:hmmformalism} for a discussion of the HMM prior). The same
band is analysed in
Ref.~\citep{2017lscHMM}. For
computational convenience, we split the band
into blocks of
approximately 0.61 Hz
(discussed further in Section~\ref{ssec:workflow}).

The final electromagnetically measured parameter is the polarization angle,
$\psi$. Because the \fstat{} components of the \jstat{} are maximized over the
polarization angle, the \jstat{} is insensitive to $\psi$.

A summary of the
search ranges flowing from the
electromagnetically measured
parameters of Sco~X-1 is presented in
Table~\ref{tbl:scox1params}.

\begin{table*}
	\caption{Electromagnetic measurements of the sky position and binary
	orbital parameters of Sco~X-1. The uncertainties
	represent one sigma (68\%) confidence intervals, except for $a_0$, for
	which hard limits are given. The search resolution for $a_0$ and $\Tasc$
	is different in each frequency sub-band, as discussed in
	Section~\ref{ssec:scox1params}. The search range for the time of ascension
	is the observed time of ascension, propagated forward to the start of O2.
	}
	\label{tbl:scox1params}
	\begin{tabular}{llllll}
		\hline
		Observed parameter & Symbol & Value & Reference
		\\ \hline
		Right ascension & $\alpha$ & $16\,\mathrm{h}\ 19\,\mathrm{m}\ 55.0850\,\mathrm{s}$ & \citep{1999bradshaw} \\
		Declination & $\delta$ & $-15^\circ38^\prime24.9^{\prime\prime}$ & \citep{1999bradshaw} \\
		Orbital period & $P$ & $68\,023.86048 \pm 0.0432$ s & \citep{WANG_NEW_SCOX1}  \\
		Projected semi-major axis & $a_0$ & $[1.45, 3.25]$ s & \citep{WANG_NEW_SCOX1} \\
		Polarisation angle & $\psi$ & $234 \pm 3^\circ$ & \citep{2001fomalont} \\
		Orbital inclination angle & $i$ & $44 \pm 6^\circ$ & \citep{2001fomalont} \\
		Time of ascension & $\Tasc$ & $974\,416\,624 \pm 50\,\mathrm{s}$ & \cite{WANG_NEW_SCOX1,2015messenger} \\
		\hline
		Search parameter & Symbol & Search range & Resolution \\
		\hline
		Frequency & $f_0$ & 60 -- 650 Hz & $5.787037 \times 10^{-7}$ Hz \\
		Projected semi-major axis & $a_0$ & 1.450 -- 3.250 s & variable \\
		Time of ascension & $\Tasc$ & $1\,164\,543\,014$ -- $1\,164\,543\,614$ s & variable \\
		\hline
	\end{tabular}
\end{table*}

\begin{figure}
	\begin{center}
	\includegraphics{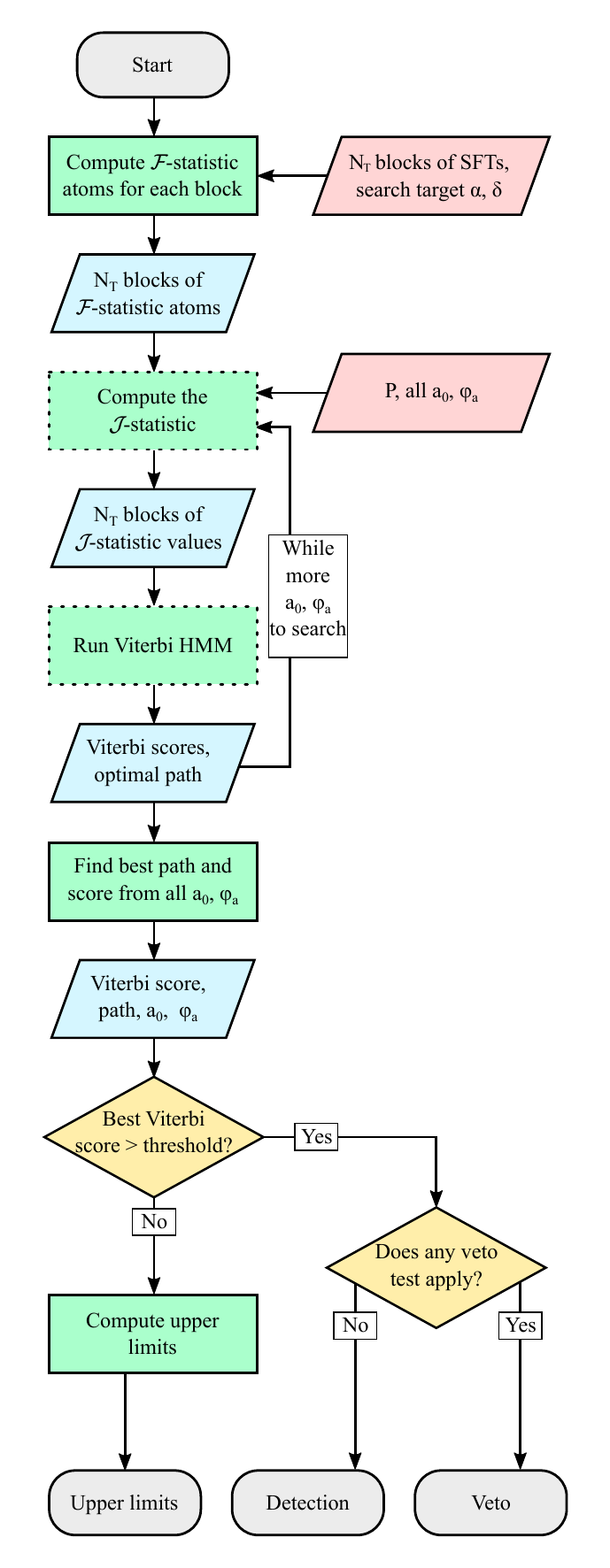}
	\end{center}
	\caption{Flowchart of the \jstat{}
	search pipeline for each sub-band. Note that the
	\fstat{} atoms are computed once per block and per sub-band, then
	the \jstat{} is recalculated for each $(a_0, \phi_a)$ pair. The grey ovals
	are the start and end of the algorithm, the green rectangles are
	procedures, the blue (red) parallelograms are intermediate
	(input) data,
	the yellow
	diamonds are decision points, and the grey dashed line represents a loop
	repeated once for each choice of parameter. The rectangles with a dashed
	boundary were run on graphical processing units (GPUs), while those with a solid boundary were run on
	central processing units (CPUs).}
	\label{fig:flowchart}
\end{figure}
\subsection{Workflow}
\label{ssec:workflow}
The workflow for the search is displayed as a flowchart in
Figure~\ref{fig:flowchart}.

The data from the detector are provided as short Fourier
transforms (SFTs), each covering $\Tsft = 1800\,\mathrm{s}$.
We divide the search into sub-bands, both to facilitate managing the volume of
data, and to ensure that replacing the search frequency $f$ with the
mid-point of the sub-band, $\bar{f}$, is a good approximation in
Equation~(\ref{eqn:convolution}).
To achieve best performance from the fast Fourier transforms used to compute
the convolution in (\ref{eqn:convolution}), it is desirable to have a
power of two number of frequency bins in the band, so we set the sub-band
width to be
$\fband = 2^{20} \fdrift =
0.6068148\,\mathrm{Hz}$.
This in turn sets the number of
hidden states
per sub-band per binary orbital parameter to be
$N_Q
= 2^{20}$.

For each sub-band, we divide the data
into $N_T$ blocks, each with duration $\Tdrift =
10\,\mathrm{d}$.
We then compute, from the SFTs, the \fstat{} ``atoms''
\citep{prixCFStechnicalnote} ($\mathcal{F}_a$, $\mathcal{F}_b$)
for each block using the fixed
parameters ($\alpha$, $\delta$, $P$) in Table~\ref{tbl:scox1params}.

The next step is to compute the \jstat{} for the ($a_0$, $\phi_a$)
search grid described in
Section~\ref{ssec:scox1params}.
The \fstat{} atoms do not depend on the binary orbital parameters so they
are not recomputed when calculating the \jstat{}.
The code to compute the \jstat{} is based on the \fstat{}
subroutines
contained in the LIGO Scientific Collaboration (LSC) Algorithm Library
(LAL) \citep{lalsoft}.

After computing the \jstat{}, we use the Viterbi algorithm to compute the
optimal paths through the HMM trellis,
i.e. the set of vectors $Q^\star$.
In principle,
the tracking problem
is three-dimensional
(over $f_0$, $a_0$ and $\phi_a$), but $a_0$ does not vary significantly
over $\Tobs \lesssim 1\,\mathrm{yr}$
and $\phi_a$ varies deterministically,
with the phase at timestep $n$ given by
$\phi_a(t_n) = \phi_a(t_{n-1}) + 2\pi\Tdrift/P$.
Thus, it is convenient to
search independently over $f_0$ and pairs $(a_0, \phi_a)$.
This allows searches over $(a_0, \phi_a)$ pairs to be performed in
parallel.

The result of this procedure
is one log likelihood for the optimal path through the trellis
terminating at every 3-tuple $(f_0, a_0, \phi_a)$.
Equation~(\ref{eqn:viterbiscore}) converts these log likelihoods
to Viterbi scores. As the noise power spectral density (PSD) of the detector
is a function of $f_0$, we compute $\mu$ and $\sigma$ separately for each band.
By contrast, the
PSD is not a function of $a_0$ and $\phi_a$. Therefore, we can
recalculate $\mu$ and $\sigma$ for every $(a_0, \phi_a)$ pair (rather than
calculating $\mu$ and $\sigma$ using every log likelihood across the entire
search), thereby considerably reducing memory use.
This has
no significant impact on the Viterbi scores.

For each sub-band that produces a best Viterbi score lower than the
detection threshold
(chosen in Section~\ref{ssec:threshold}), we compute an upper limit on the
gravitational wave strain for a source in that sub-band. For
Viterbi scores that exceed the threshold, we apply the veto tests described in
Section~\ref{sec:veto}. We claim a detection, if a candidate survives all
vetoes.

For performance reasons, the most computationally-intensive parts of the search
(computing the \jstat{}, and the Viterbi tracking) were run using NVIDIA P100 graphical
processing units (GPUs). Other steps were run using CPU codes, on Intel Xeon
Gold 6140 CPUs.

\subsection{Threshold and false alarm probability}
\label{ssec:threshold}

\begin{figure}
	\includegraphics[width=0.49\textwidth]{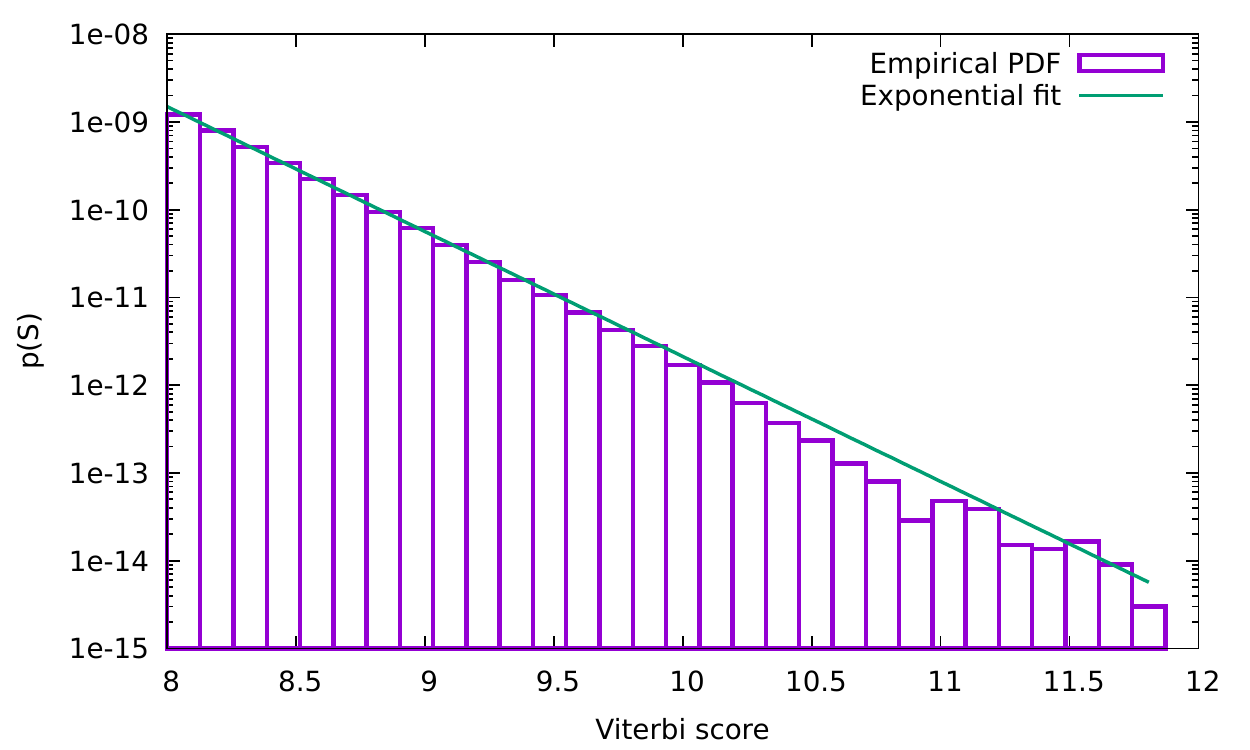}
	\caption{Tail of the PDF of the Viterbi score $S$ in noise. The purple histogram shows
	the empirical PDF derived from $10^2$ realisations of the noise, analysed
	in the seven 0.61 Hz sub-bands starting at
	55, 155, 255, 355, 455, 555 and 650 Hz. The green curve
	is an exponential fitted to the histogram.}
	\label{fig:threshold}
\end{figure}

It remains to determine a detection score threshold $\Sth$
corresponding to the desired false alarm probability.
Consider the probability density function (PDF) $p_{\mathrm{n}}(S)$ of the Viterbi
score in noise. For a given threshold $\Sth$ and a fixed search frequency and
set of binary orbital parameters, the probability that the score
will exceed this threshold (i.e. produce a false alarm) is
\begin{align}
	\alpha = \int_{\Sth}^{\infty} \mathrm{d}S\, p_{\mathrm{n}}(S) .
	\label{eqn:farPDF}
\end{align}
In general, the search covers many frequency bins and choices of binary
parameters. The probability $\alpha_N$ of a false alarm over a search covering $N$
parameter choices (number of frequency bins multiplied by number of binary
parameter choices) is
\begin{align}
	\alpha_N = 1 - (1 - \alpha)^N.
	\label{eqn:farEqn}
\end{align}
This equation assumes that the Viterbi score in noise is an independent random
variable at each point in the parameter space,
which is not necessarily true, as the \jstat{} calculated for two
points nearby in parameter space
are correlated to some degree. However, for $\mu_{\mathrm{max}} = 0.1$ as used in this
search, these correlations do not have a significant impact \citep{2012wette}.
In practice, we fix $\alpha_N$ and $N$ and solve (\ref{eqn:farPDF}) and
(\ref{eqn:farEqn}) for $\alpha$ and hence $\Sth$.

As the noise-only PDF $p_{\mathrm{n}}(S)$ of the Viterbi score is unknown analytically
\citep{2017lscHMM}, we resort
to Monte-Carlo simulations.
We generate $10^2$ Gaussian noise realisations in
seven sub-bands of width $\fband$, namely those
starting at 55~Hz, 155~Hz, 255~Hz, 355~Hz, 455~Hz, 555~Hz and 650~Hz.
The noise is generated using the
standard LIGO tool \textit{lalapps\_Makefakedata\_v4}.
These are the same sub-bands used in Section~IIIC of Ref.~\citep{2017lscHMM},
and
the one-sided noise PSD $S_h(f)$ is set to match the
O2 data.
We then perform the search described in Section~\ref{ssec:workflow}
(including scanning over $a_0$ and $\phi_a$).

The results of this search produce an empirical
version of $p_{\mathrm{n}}(S)$.
Plotting the tail of this distribution on a logarithmic plot suggests
that a fit to a function of the form $e^{\lambda S}$
is an appropriate choice to allow the PDF to be
extrapolated in order to solve (\ref{eqn:farEqn}).

We first analyse each band independently, to ensure that there is no frequency
dependence in $p_{\mathrm{n}}(S)$.
Table~\ref{tbl:far} gives the best-fit $\lambda$, and the threshold $\Sth$
obtained, for each
band
analysed in isolation.
We find that there is no significant dependence on the sub-band searched, nor
any identifiable trend in $\lambda$ or $\Sth$.
Combining the realisations for all bands produces $\lambda = -3.28$ and
hence $\Sth = 13.66$ for $\alpha = 0.01$.
The empirical PDF and fitted exponential are shown in
Figure~\ref{fig:threshold}.

\begin{table}
	\caption{Results of investigating the empirical PDF of the Viterbi
	score in seven sub-bands in Gaussian noise.
	The second column is $\lambda$ obtained by fitting the PDF to $e^{\lambda
	S}$. The third
	column is the threshold $\Sth$ obtained by solving equation
	(\ref{eqn:farEqn}).
	}
	\label{tbl:far}
	\begin{center}
	\begin{tabular}{ccc}
		\hline
		Start of band (Hz) & $\lambda$ & $\Sth$
		\\ \hline
		$55$  & $-3.02$  & $14.12$ \\
		$155$ & $-3.24$  & $13.63$ \\
		$255$ & $-3.26$  & $13.58$ \\
		$355$ & $-3.27$  & $13.61$ \\
		$455$ & $-3.30$  & $13.62$ \\
		$555$ & $-3.29$  & $13.66$ \\
		$650$ & $-3.29$  & $13.63$ \\
		\hline
	\end{tabular}
	\end{center}
\end{table}

\subsection{Sensitivity}
\label{ssec:sensitivity}

\begin{figure*}
	\includegraphics[width=\textwidth]{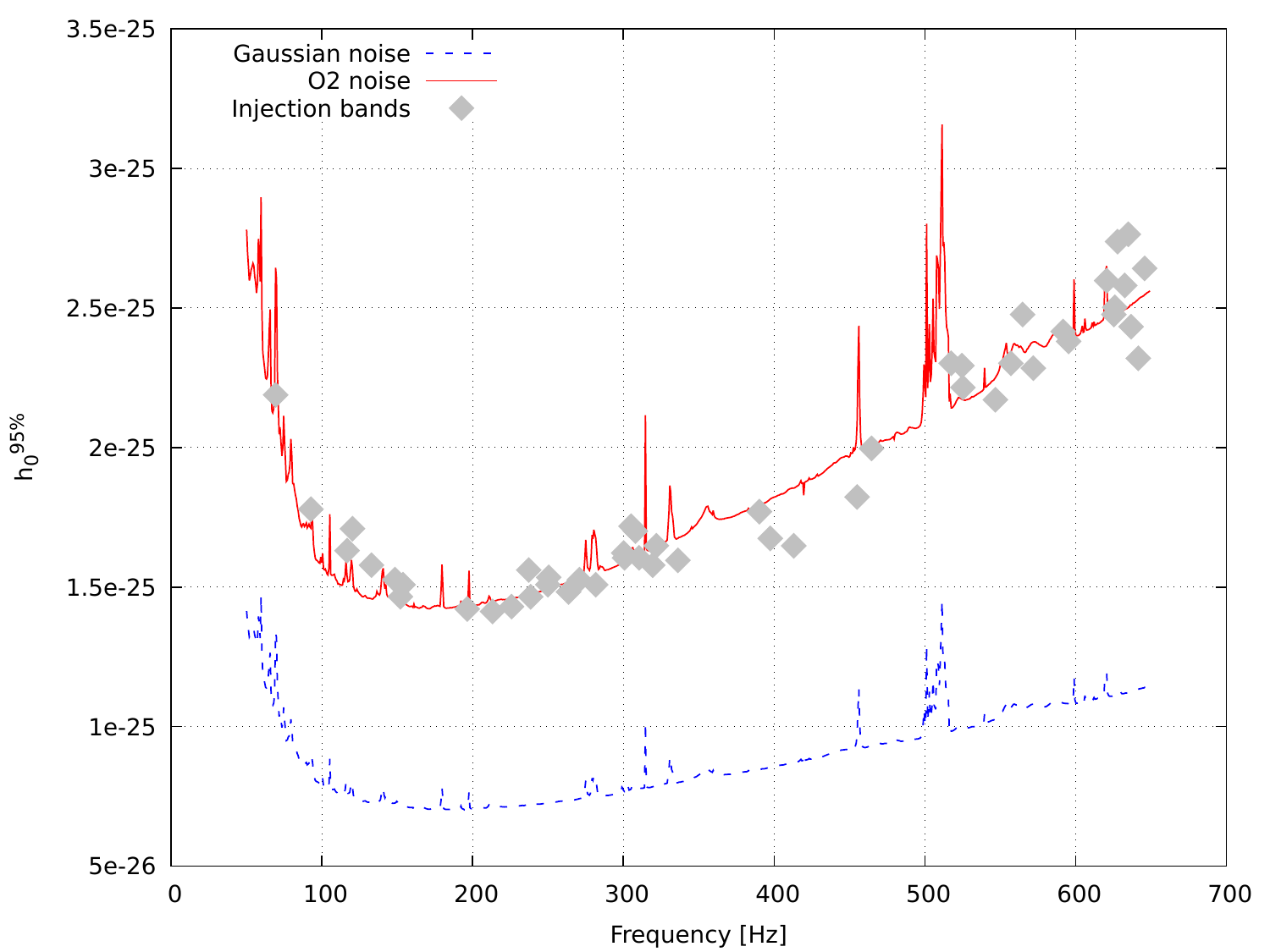}
	\caption{Sensitivity of a search for Sco~X-1, across the frequency
	band searched in this work.
	The horizontal axis shows the search frequency. The vertical axis
	shows the wave strain $h_0^{95\%}$ needed for a 95\% detection
	efficiency, on the assumption $\cos\iota = 1$, namely, a circularly polarized signal.
	The blue
	dashed curve is based on simulations of Gaussian noise, while the red solid
	curve is corrected for the non-Gaussian statistics of the noise and the
	interferometer duty cycle, through multiplying by
	$\kappa_{\mathrm{freq}}(f)$
	(see Section~\ref{ssec:sensitivity}). The
	diamonds show $h_0^{95\%}$ derived through injections into sub-bands, again
	assuming a circularly polarized signal (see
	Section~\ref{ssec:sensitivity}).
	}
	\label{fig:h95}
\end{figure*}

After selecting $\Sth$, it remains to determine the
lowest (as a function of frequency)
characteristic wave strain, $h_0^{95\%}$, that can be detected with 95 per cent
efficiency (i.e., a five per cent false
dismissal rate). To do this, we generate Monte-Carlo realisations of Gaussian
noise with Sco~X-1--like signals injected.
We determine the proportion of signals recovered
as a function of $h_0$
and double-check the false alarm probability
quoted above.

For O2, the most sensitive sub-band
of width $\fband = 0.6068148\,\mathrm{Hz}$
is the one beginning at 194.6 Hz.
Following a typical procedure used to find upper limits for continuous
gravitational wave searches \citep{ScoX1S2paper},
we generate $10^2$ noise realisations and inject
signals, using the source parameters
in Table~\ref{tbl:scox1params},
with
$\Tobs = 230\,\mathrm{d}$ (the duration of O2),
$\Tdrift = 10\,\mathrm{d}$,
$N_T = 23$,
$\sqrt{S_h} = 7.058 \times 10^{-24} \mathrm{Hz}^{-1/2}$, and $\cos\iota = 1$.
The remaining range-bound parameters, namely
$f_{0_{\mathrm{inj}}}$, $a_{0_{\mathrm{inj}}}$,
$T_{{\mathrm{asc}}_\mathrm{inj}}$ and
$\psi_{\mathrm{inj}}$ are chosen from a uniform distribution within the range
given by their 1-$\sigma$ error bars. The source frequency
$f_{0_{\mathrm{inj}}}$ is chosen from a uniform distribution on the interval
$[194.6\,\mathrm{Hz}, 194.7\,\mathrm{Hz}]$.
For each realisation, the signal is injected with progressively lower $h_0$
until it can no longer be detected. We denote by $h_{0,\mathrm{min};i}$ the
lowest $h_0$ that can be detected in realisation $i$.
To obtain $h_0^{95\%}$, we take the 95th highest $h_{0,\mathrm{min};i}$.
The simulations return the threshold
$h_0^{95\%} = 1.46 \times 10^{-25}$ at 194.6 Hz.

In general, the signal-to-noise ratio is strongly
affected by the inclination angle $\iota$,
not just $h_0$.
We
follow Ref.~\citep{2015messenger} and define an effective $h_0$ that absorbs
the dependence on $\iota$:
\begin{align}
	\heff = h_0 2^{-1/2}\{ [ (1 + \cos^2 \iota)/2 ]^2 + \cos^2\iota \}^{1/2},
	\label{eqn:heff}
\end{align}
allowing us to generalize results from the simulations above,
where all injections were done with $\cos\iota =
1$.
Thus,
the result obtained above
corresponds to
circular polarization.
The
electromagnetically measured inclination of of Sco~X-1's orbit is 
$i \approx 44^\circ \pm 6^\circ$ \citep{2001fomalont}. Although it is not necessarily the
case, if we assume that the orbital
inclination equals the inclination angle $\iota$ of the putative neutron
star's spin axis, we obtain $h_0^{\mathrm{\iota\approx44^\circ,95\%}} = 1.35
h_0^{\mathrm{eff,95\%}}$.

The search in Ref.~\citep{2017lscHMM} found a scaling relation of the form
	$h_0^{95\%} \propto S_h^{1/2}f_0^{1/4}$,
to hold
for fixed $\Tobs$. The $f_0^{1/4}$ dependence arises because the latter
search
added sidebands incoherently. In the case of the \jstat{}, which adds sidebands
coherently, we expect the scaling to depend just on $h_0$, with
\begin{align}
	h_0^{95\%} \propto S_h^{1/2}.
	\label{eqn:snrfSh}
\end{align}
We verify this scaling in Gaussian noise by repeating the injection procedure
described above in frequency bands beginning at 55 Hz, 355 Hz and 650 Hz.
The scaling is the final ingredient needed to produce the blue
dashed curve in Figure~\ref{fig:h95}, which shows the expected sensitivity of a
search over the full search band, assuming Gaussian noise, a 100 per cent
duty cycle and a circularly polarized signal.

There is no simple scaling similar to (\ref{eqn:snrfSh})
that can be used to account for the effect of non-Gaussian noise and the
detector duty cycle.
Hence we
introduce a multiplicative correction
factor $\kappa_j$ for a selection of sub-bands indexed by $j$, following Ref.~\citep{2017lscHMM}.
We determine
$\kappa_j$ by doing $10^2$ injections
(drawing parameters as described above)
into the detector data for the $j$-th sub-band, again using progressively-lower
$h_0$ until we determine the minimum $h_0$ detected.
Then,
$\kappa_j$ equals $h_0^{\mathrm{eff,95\%}}$ for injections into real
noise, divided by $h_0^{\mathrm{eff,95\%}}$ for injections into Gaussian noise.

Producing $\kappa_j$ in this way for a random selection of sub-bands in the
search band suggests that $\kappa$ depends weakly on frequency, most likely due
to the \jstat{} not perfectly summing sidebands \citep{2017lscHMM}. A linear fit to the computed
$\kappa_j$ values suggests a frequency-dependent correction factor
\begin{align}
	\kappa_{\mathrm{freq}}(f) = 1.944 + 4.60\times10^{-4} f / (1\, \mathrm{Hz}).
\end{align}
We use $\kappa_{\mathrm{freq}}(f)$ to adjust the blue dashed curve in
Figure~\ref{fig:h95}, producing the red solid curve in that figure, which
represents the expected sensitivity across the full search band, where the
noise is realistic (i.e. not Gaussian). The 50 sub-bands sampled are
shown on the plot as grey diamonds.

\section{O2 analysis}
\label{sec:analysis}

\begin{figure*}
	\begin{subfigure}[b]{0.49\textwidth}
		\includegraphics[width=\textwidth]{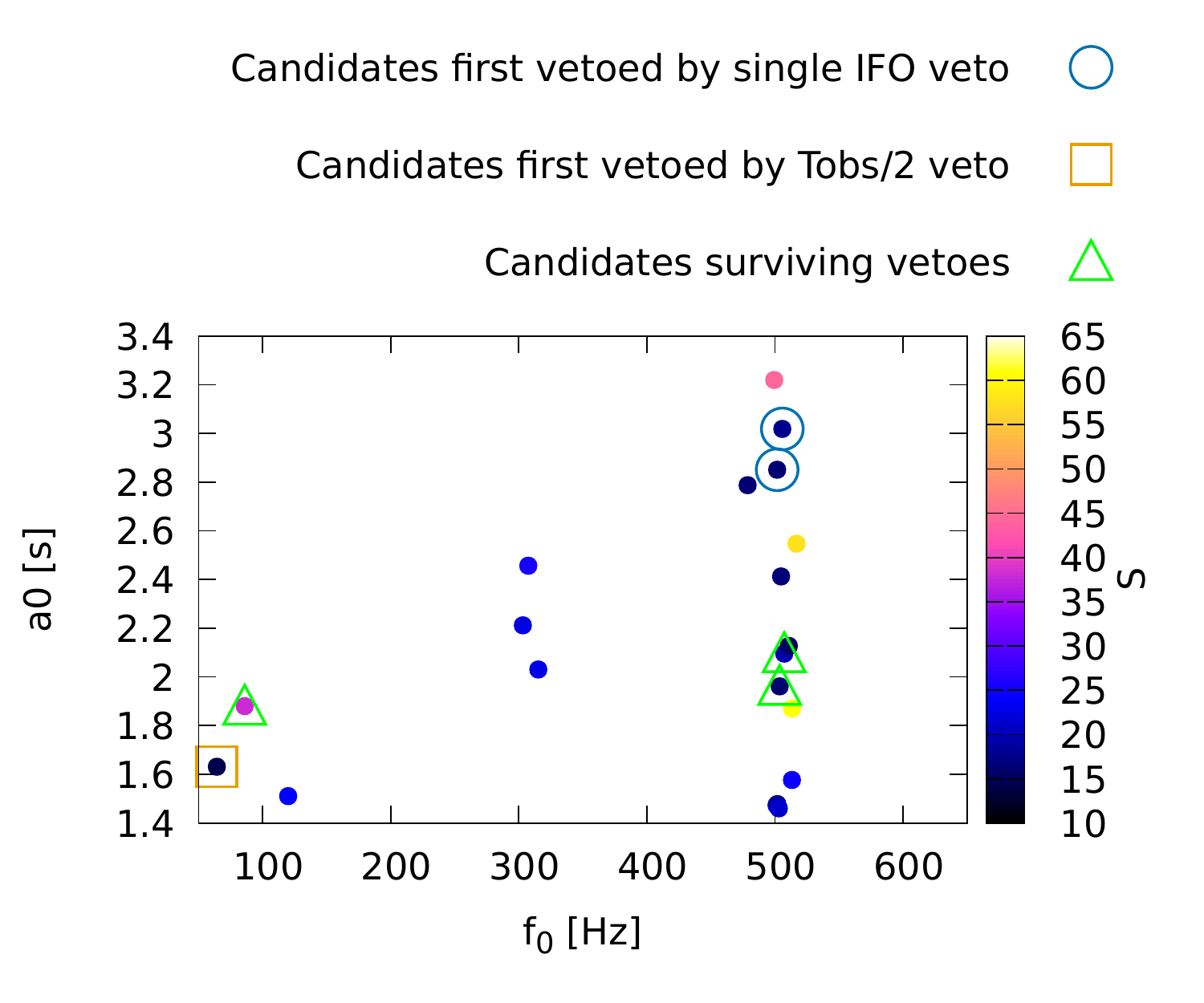}
	\end{subfigure}
	\hfill
	\begin{subfigure}[b]{0.49\textwidth}
		\includegraphics[width=\textwidth]{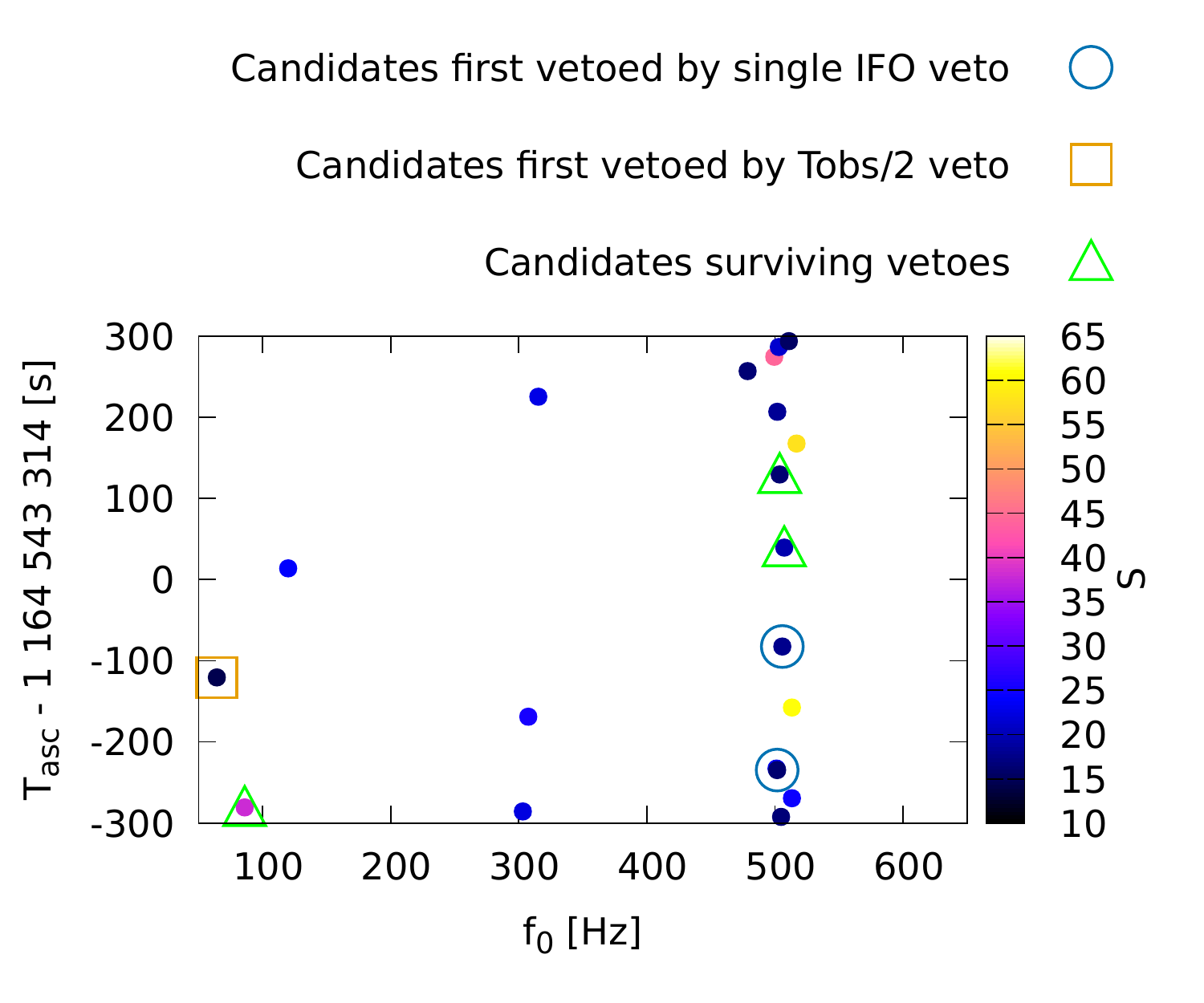}
	\end{subfigure}
	\caption{Candidates identified by the search. The left hand panel plots the
	detection score $S$ (indicated by colour; see colour bar)
	as a function of final frequency
	$f_0(t_{N_T})$ (horizontal axis) and orbital
	semi-major axis $a_0$
	(vertical axis) recovered by the HMM.
	The right hand panel plots the candidates with $\Tasc$ on the vertical
	axis.
	Undecorated candidates
	are eliminated by the known line veto, candidates marked by blue circles are
	eliminated by the single interferometer veto, candidates marked by
	orange squares
	are eliminated by the $\Tobs/2$ veto, and the candidates marked by green
	triangle survive the veto process.}
	\label{fig:result}
\end{figure*}

We now analyse the data from LIGO's Observing Run 2 (O2), using the full dataset
from 30 November 2016 to 26 August 2017, including data from the LIGO Livingston
(L1) and Hanford (H1) observatories.
The Virgo interferometer also participated
in the last two months of O2, but we do not use any Virgo data in this
analysis.

There are two notable pauses
in data gathering: an end-of-year break starting on 22 December 2016 lasting
for 13 days,
and a commissioning break starting on 7 May 2017 lasting for 19 (L1) or 32 (H1)
days.

Data stretches
shorter than $\Tsft$ are discarded, as is
a period of approximately
one month where much of the band was contaminated,
due to a blinking light in the power system and a
digital camera (used for detector diagnostics) that was inadvertantly left on.
A detailed discussion of Advanced LIGO detector noise can be found in
Ref.~\citep{2018covas}.
Taking all these factors into account, the overall duty cycle (i.e. proportion
of time spent gathering science-quality
data) for O2 was 51.9\% (L1) and 46.2\% (H1).

Because of the commissioning break, 
one ten-day block has no data.
We fill this block with uniform log likelihood,
so that
the HMM has no
preference for remaining in the same frequency bin, or moving by one bin,
during the break,
while still allowing a maximum drift of $\fdrift$ every ten days.
An alternative, but equivalent, approach would be to remove the break entirely,
and alter the
transition matrix $A_{q_iq_j}$ for that step to allow the HMM to wander up to
two frequency bins.
The end-of-year break is also longer than ten days,
but it is covered by two blocks.
Both of the blocks that overlap with the end-of-year
break contain data.

We search the same frequency band as Ref.~\citep{2017lscHMM}, namely 60 --
650 Hz. The lower limit is set by LIGO's poor sensitivity for signals $\lesssim
25\,\mathrm{Hz}$ and the significant contamination from instrumental noise in
the band 25 -- 60 Hz. The sensitivity of the search falls as frequency
increases, while compute time rises dramatically.
We terminate the
search at 650 Hz, as in Ref.~\citep{2017lscHMM}.

The results of the search are presented in Figure~\ref{fig:result}, which shows the
frequency and recovered orbital parameters $a_0$ and $\phi_a$ for every
path with $S > \Sth$.
The colour of the points shows the Viterbi score associated with that path.
As the most a signal can wander during the
observation is $N_T \fdrift \approx 1.3 \times 10^{-5}\,\mathrm{Hz}$, which
is
small compared to $\fband$ (and what can be visually discerned on
Figure~\ref{fig:result}), we define $f_0$ for a given path to be equal to
$f_0(t = N_T)$ for convenience.

To rule out false alarms,
we apply the
hierarchy of vetoes first described in Ref.~\citep{2017lscHMM}.
The vetoes are: (1) the known instrumental lines veto (described in
Section~\ref{ssec:knownlines} below), (2) the single interferometer veto
(Section~\ref{ssec:singleifo}), (3) the $\Tobs/2$ veto
(Section~\ref{ssec:tobs}) and (4) the $\Tdrift$ veto
(ultimately not used, but discussed in Section~IV\,A\,4 of Ref.~\citep{2017lscHMM}).
To ensure that the vetoes are unlikely to
falsely dismiss a true signal, we perform the search on a dataset with
synthetic signals injected into it, and ensure that those injections are not
vetoed.
These veto safety tests are
described in Section~\ref{sec:vetosafe}.

The number of candidates found in the initial search, and then vetoed at each
step, are listed in Table~\ref{tbl:results}.

\subsection{Vetoes}
\label{sec:veto}

\subsubsection{Known lines veto}
\label{ssec:knownlines}
There are a large number of persistent
instrumental noise lines identified as part of
LIGO's detector characterisation process \citep{2017lscStochastic,2018covas}.
These lines can arise from a number of sources, including interference from
equipment around the detector, resonant modes in the suspension
system, and external environmental causes (e.g. the electricity grid).

A noise line generally produces high
$|\mathcal{F}_a|$ and $|\mathcal{F}_b|$
values. The convolution in (\ref{eqn:convolution}) reduces the impact
of this somewhat by summing bins near and far from the line%
, but in practice the noise lines are strong enough that they contaminate
any candidate nearby.
Accordingly, we veto any candidate whose Viterbi path $f_0(t)$ satisfies $\left| f_0(t) - \fline \right| <
2\pi a_0 f_0 / P$, for any time $t$ along the path and for any line frequency
$\fline$. This veto is efficient, excluding 14 of the 20
candidates.

\subsubsection{Single interferometer veto}
\label{ssec:singleifo}
During O2, L1 was slightly more sensitive than H1, but overall the
sensitivities of the two interferometers were similar. Accordingly, any astrophysical
signal that can be detected in the combined dataset should either be detected
by
the individual detector datasets when analysed separately (for stronger
signals) or in neither (for weaker signals). A signal
that is detectable in one interferometer only
is likely to be a noise artifact, so we veto it.

Following Ref.~\citep{2017lscHMM}, we compare the Viterbi scores obtained from
individual detectors to the original combined score $S_\cup$ to
classify survivors of the known line veto into
four categories, discussed below, one of which is vetoed.

\textit{Category A}. One detector returns $S < \Sth$, while the
other detector returns $S > S_\cup$, and the frequency estimated by the latter
detector is close to that of the original candidate
$f_{0\cup}$, that is, $|f_{0\cup} - f_0| < 2\pi a_{0\cup} f_{0\cup} /
P$, where
the subscript $\cup$ denotes a quantity
estimated by the search in both detectors. This category, and the next,
represent signals where the score is dominated by one detector.
We veto candidates in Category A.

\textit{Category B}. As with Category A,
one detector returns
$S < \Sth$,
while the other detector returns $S > S_\cup$. Unlike Category A, the
frequency estimated by the latter detector
is far from the original candidate, i.e., $|f_{0\cup} - f_0| > 2\pi a_{0\cup}
f_{0\cup} / P$.
In this case, it is possible that there is signal at $f_{0\cup}$ which is
detectable
when combining the data from both detectors but not from one detector, because
some artifact
masks its presence.
Hence we keep the candidate
for follow-up.

\textit{Category C}. The candidate is seen with $S > \Sth$ in both detectors.
This could either be a relatively strong signal, or an artifact from a
noise source common to both detectors. The single interferometer veto cannot
distinguish these possiblities. Again, we keep the candidate for follow-up.

\textit{Category D}. The candidate is not seen by either detector, with $S <
\Sth$ in both detectors. This could be a signal that is too weak to see in either detector
individually. We keep the candidate for follow-up.

Category A of the single-interferometer
veto eliminates two of the remaining six candidates.
The two eliminated candidates were stronger in H1 compared to L1.

\subsubsection{$\Tobs/2$ veto}
\label{ssec:tobs}
We divide the observing run into two segments, the first covering 140 days from
30 Nov 2016 (GPS timestamp $1\,164\,562\,334$) to 19 Apr 2017 (GPS timestamp
$1\,176\,658\,334$), and the second covering 90 days from 19 Jan 2017 (GPS
timestamp $1\,168\,882\,334$) to 25 Aug 2017 (GPS timestamp $1\,187\,731\,792$).
This division is chosen to get approximately equal effective observing time in the two
segments.
There is no forceful evidence to suggest that the
gravitational wave strength of an LMXB
varies significantly with time (and a signal with time-varying strength is
likely to have a considerably more complicated form than assumed here);
thus we do not expect a signal to
appear preferentially in either segment.
We search the segments separately for the candidates which survived
both preceeding vetoes.
To determine whether to veto candidates at this stage, we
apply the same set of categories as in veto 2.

This veto eliminates one remaining candidate, which is much stronger in the
first segment of the observing run than the second.

Ref.~\citep{2017lscHMM} describes the $\Tdrift$ veto as a fourth veto that can
be applied to candidates surviving the $\Tobs/2$ veto. However, this veto is
applicable to candidates with an observed spin wandering timescale that is
20 days or longer. This is not the case for the surviving three candidates, so
the $\Tdrift$ veto is not applicable to them.

The remaining candidates are in the sub-bands starting at 85.4 Hz, 503.6 Hz and
507.2 Hz. The scores relevant to performing the veto procedure are given in
Table~\ref{tbl:unvetoed}.
All three candidates are stronger when analyzing the H1 detector data alone
compared to analyzing L1 detector data alone, with the L1 results consistent
with noise.
The candidates in the sub-bands
starting at 85.4 Hz and 507.2 Hz are both stronger during the second half of O2
compared to the first half, while the candidate in the sub-band starting at
503.6 Hz is stronger in the analysis of the first half of O2.
Particularly for the candidate in the 85.4 Hz sub-band, the asymmetry
in score between the first and second half of the observing is extreme and
suggestive of a detector artifact rather than an astrophysical signal.
The asymmetry is less prononced for the candidates in the sub-bands starting at
503.6 Hz and 507.2 Hz, but both of these candidates are
in a region of frequency space that is significantly contaminated
by interferometer
noise, particularly violin modes associated with the LIGO mirror suspension.
For these reasons, it is most likely that these candidates are due to unknown
instrumental
noise in the H1 detector, although they are not formally ruled out by the veto
procedure described above.

\begin{table*}
	\caption{
Viterbi scores of the three candidates that survived the veto procedure. The
	original score is the score of the original candidate, from the search on
	the full O2 dataset. The H1 and L1 scores are the scores for the candidate
	when searching on each detector independently. The first and second part
	scores are the scores when analyzing the first 140 and last 90 days of the
	dataset, respectively.
		}
	\label{tbl:unvetoed}
	\begin{tabular}{l|l|ll|ll}
		Sub-band containing candidate & Original score & H1 & L1 & First part &
		Second part \\
		\hline
		85.4  & 42.4 & 30.7 & 6.3 & 7.2  & 41.8  \\
		503.6 & 41.3 & 34.6 & 5.8 & 37.5 & 6.1  \\
		507.2 & 17.3 & 10.6 & 6.1 & 10.2 & 16.4 \\
	\end{tabular}
\end{table*}

\subsection{Veto safety}
\label{sec:vetosafe}
To verify that the vetoes described previously do not unduly increase the false
dismissal probability, we inject signals into the O2 data and perform the veto
procedure described in the previous section.
We inject a total of 50 signals
into 50 sub-bands of width $\fband$, chosen to be comparable to the 200
injections used for the equivalent tests in Ref.~\citep{2017lscHMM} while
having a large enough sample to be confident that false dismissals caused by the
vetoes are rare in the context of the five per cent false dismissal rate used
in calculating sensitivity.
The sub-bands and parameters chosen are selected randomly from the
search band to achieve good frequency coverage,
but excluding those sub-bands that contain a known line (and hence
would be excluded by the known lines veto).
Into
these sub-bands, we inject a signal near the detection limit
with $h_0$ typically at $h_0^{95\%}$ for that sub-band (although we inject a
stronger signal if the signal turns out to be undetectable), and
with $f_0$ drawn randomly from a uniform distribution over the interval
$[f_{\mathrm{start}} + 0.1\,\mathrm{Hz},
f_{\mathrm{start}} + \fband - 0.1\,\mathrm{Hz}]$, where $f_{\mathrm{start}}$
is the lowest frequency in the sub-band. At each block, the signal is allowed
to wander at most one frequency bin (i.e., by an amount drawn uniformly from
$[-\fdrift, +\fdrift]$), and the signal frequency is constant within the block,
following Ref.~\citep{2017lscHMM}.
The other parameters chosen in the same way as for
the sensitivity tests described in Section~\ref{ssec:sensitivity}.

We then apply vetoes 2 (single interferometer veto) and 3 ($\Tobs/2$ veto)
to each candidate (veto 1 is inapplicable,
as the injection bands avoid known lines; and veto 4 [$\Tdrift$ veto] was not
used in this
search). No injection was vetoed.

Because the veto safety procedure uses the O2 data as noise, it is possible
that the safety results described above depend in some way on the specifics of
O2. However, as the veto procedure copies
the equivalent procedure in
Ref.~\citep{2017lscHMM}, which tests both S5 noise and O1 noise, we have
confidence that the veto safety result is not specific to the peculiarities of
O2.

\begin{table}
	\caption{Number of candidates found in the first pass,
	and number remaining after applying the vetoes described in
	Section~\ref{sec:veto}.}
	\label{tbl:results}
	\begin{tabular}{lr}
		\hline
		After veto & Survivors \\
		\hline
		First pass      & 20 \\
		Line & 6 \\
		Single interferometer & 4 \\
		$\Tobs/2$ & 3 \\
		\hline
	\end{tabular}
\end{table}

\section{Upper limits}
We can use the non-detection reported in the previous section, in concert with
the approach outlined in
Section~\ref{ssec:sensitivity}, to place an
upper limit on $h_0$ as a function of $f_0$ and compare the result to the
indirect,
torque-balance upper limit established by the X-ray flux \citep{1998bildsten}.

\subsection{Frequentist upper limit at 95\% confidence}

\begin{figure*}
	\includegraphics[width=\textwidth]{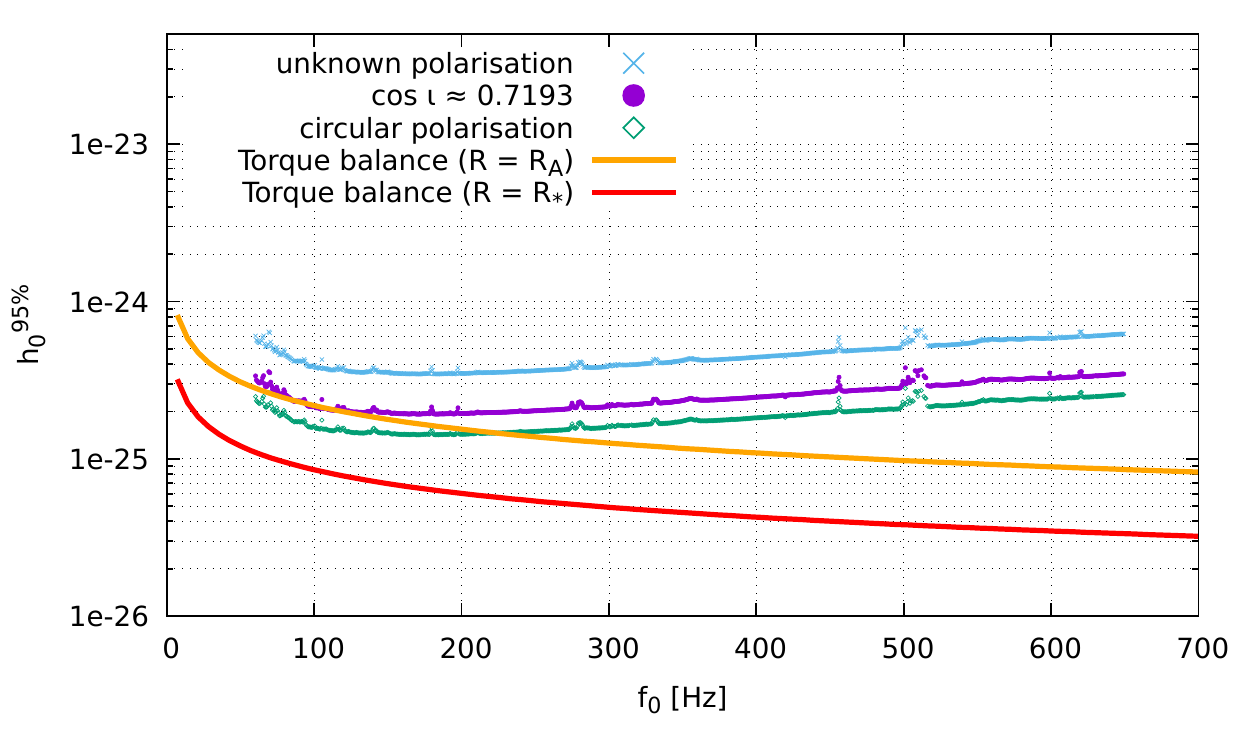}
	\caption{Wave strain upper limits at 95\% confidence as a function of
	signal frequency,
	corresponding to
	the
	frequentist upper limit in each sub-band (width $\fband$),
	for three
	scenarios:
	a flat prior on $\cos\iota$ (blue crosses),
	the orientation $\iota \approx 44^\circ$ derived from observations of the
	radio jet (filled purple circles)
	and
	the most optimistic case of circularly polarized waves (open green diamonds).
	These are compared to the
	indirect torque-balance upper limit,
	where the
	accretion torque is applied at
	the Alfv\'{e}n radius (solid orange curve) or
	the stellar radius (solid red curve).}
	\label{fig:upperlimit}
\end{figure*}

Failure to detect a gravitational wave signal allows us to place an upper limit
on $h_0$ from a particular source, given a desired confidence
level. In this section, we follow Ref.~\citep{2017lscHMM} in using a frequentist
approach
and setting
95\%
as the desired confidence level.
The alternative,
Bayesian approach in Ref.~\citep{2015lscSidebandScoX1}
is hard
to adapt to the HMM-based search, because correlations between the Viterbi paths
render the distribution of Viterbi scores difficult to calculate analytically.

We define $h_0^{95\%}$ to be the lowest amplitude signal for
which we have a 95\% probability
or
greater of detecting a signal with $h_0 \geq h_0^{95\%}$,
that is, $\mathrm{Pr}(S
\geq \Sth | h_0 \geq h_0^{95\%}) \geq 0.95$.
The value of $\hnfive$ depends on the inclination angle of the source, through
equation~(\ref{eqn:heff}). Figure~\ref{fig:upperlimit} show the upper limit for
three cases: assuming the neutron star spin axis inclination angle $\iota$ is
equal to the electromagnetically-constrained orbital inclination angle $i
\approx 44^\circ$ (purple plus signs), a pure circularly-polarized signal
$|\cos\iota| = 1$ (green crosses), and a flat prior on $\cos\iota$ (blue
asterisks). For sub-bands with no candidate path with a Viterbi score above the
threshold,
we take $\hnfive$ from Figure~\ref{fig:h95} for the
circularly-polarized case, and determine $\hnfive$ for the two other
cases using equation~(\ref{eqn:heff}).
No upper limit is established for sub-bands containing a vetoed candidate
(because those bands are deemed to be contaminated by instrumental artifacts).
Accordingly those sub-bands are excluded from Figure~\ref{fig:upperlimit}.

The circularly-polarized case produces the most stringent upper limit,
reflecting the fact that $|\cos \iota| = 1$ would be the most favourable
configuration for producing gravitational waves. Conversely, assuming no
knowledge of the inclination angle (the flat prior case) produces a looser
upper limit.
The lowest upper limit for this search is in the sub-band starting at 194.6 Hz,
with upper limits of $\hnfive = 3.47\times10^{-25}$, $1.93\times10^{-25}$,
$1.42\times10^{-25}$ for the
unknown polarization, electromagnetically constrained, and circularly polarized
cases, respectively.
Previous work with the HMM, in Ref.~\citep{2017lscHMM}, found $\hnfive = 8.3\times10^{-25}$,
$4.0\times10^{-25}$, $3.0\times10^{-25}$ for those cases in its most sensitive sub-band,
starting at 106 Hz.

\subsection{Torque-balance upper limit}
An indirect upper
limit on gravitational wave strain can be
obtained from X-ray
observations.
If the spin-down torque due to
gravitational wave
emission
balances the
accretion spin-up torque, with the latter inferred from the X-ray luminosity,
one has $h_0 \geq \heq$ with
\citep{1998bildsten,1984wagoner,2014sammut}
\begin{align}
	\heq = &5.5 \times 10^{-27}
	\left( \frac{F_X}{10^{-8}\,\mathrm{erg}\, \mathrm{cm}^{-1}\,\mathrm{s}^{-1}}
	\right)^{1/2}
	\left(\frac{R}{10\,\mathrm{km}}\right)^{3/4}
	\notag \\
	& \left(\frac{1.4 M_\odot}{M_\star}\right)^{1/4}
	\left(\frac{300\,\mathrm{Hz}}{f_\star}\right)^{1/2},
\end{align}
where $F_X$ is the X-ray flux, $R$ is the length of the notional ``lever arm''
to which the accretion torque is applied, $M_\star$ is
the stellar mass and $f_\star$ is the (unknown) spin frequency.

To establish an upper limit, we take the electromagnetically measured $F_X = 4
\times 10^{-7}\,\mathrm{erg}\,\mathrm{cm}^{-2}\,\mathrm{s}^{-1}$
\citep{2008watts} of Sco~X-1, and the common fiducial neutron star mass
$M_\star = 1.4M_\odot$. The most conservative choice for the accretion torque
lever arm is the stellar radius $R_\star = 10\,\mathrm{km}$.
We plot $\heq$ as a function of frequency as the solid red curve in
Figure~\ref{fig:upperlimit}. Another physically reasonable choice of lever arm
length is the Alfv\'{e}n radius, $R_A$, i.e. the distance out to which outflowing material
co-rotates with the star's magnetic field. This is given by
\citep{1997bildsten,2017lscHMM}
\begin{align}
	R_A =& 35\left(\frac{B_\star}{10^9\,\mathrm{G}}\right)^{4/7}
	\left(\frac{R_\star}{10\,\mathrm{km}}\right)^{12/7}
	\notag \\
	& \left(\frac{1.4 M_\odot}{M_\star}\right)^{1/7}
	\left(\frac{10^{-8} M_\odot\,\mathrm{yr}^{-1}}{\dot{M}} \right)^{2/7}
	\mathrm{km},
	\label{eqn:alfven}
\end{align}
where $B_\star$ is the polar magnetic field strength at the stellar surface,
$G$ is Newton's
gravitational constant and $\dot{M}$ is the accretion rate. The accretion rates
in LMXBs can range from the Eddington limit,
$2\times10^{-8}M_\odot\,\mathrm{yr}^{-1}$, down to about $10^{-11} M_\odot
\,\mathrm{yr}^{-1}$ \citep{2003ritter,sammutthesis}.
The magnetic fields on the neutron stars in LMXBs are
comparatively weak, lying in the range
$10^8\,\mathrm{G} \lesssim B_\star \lesssim 10^{9}\,\mathrm{G}$
\citep{1998bildsten,sammutthesis,2012patruno}. We substitute $\dot{M} =
10^{-8}M_\odot\,\mathrm{yr}^{-1}$ and $B_\star = 10^9 \,\mathrm{G}$ into
equation~(\ref{eqn:alfven}),
to maximize
$R_A$ and hence $\heq$.
The result is
plotted as the orange curve in Figure~\ref{fig:upperlimit}. Both torque balance
curves are plotted with $f_0 = 2f_\star$, i.e. an orthogonal biaxial rotor,
which is a
conventional assumption \citep{1998jks}.

At the most sensitive sub-band, starting at $f_0 = 194.6\,\mathrm{Hz}$,
the electromagnetically-constrained upper limit is a factor of about 1.2 below
(3.1 above) the torque balance for $R = R_A$ ($R = R_\star$).
The upper limits for a circularly-polarized
signal beat the $R = R_A$ torque balance upper limit
between 60 and 223 Hz; and the upper limits assuming an electromagnetically
constrained inclination angle beat the $R = R_A$ torque
balance limit
between 94 Hz and 113 Hz.

The upper limits given in Figure~\ref{fig:upperlimit} are somewhat higher than those
achieved by the most sensitive search to date, the O1 cross-correlation search,
which has upper limits that are typically lower by a factor of approximately
1.5
\citep{2017lscO1crosscorr}.
A significant contributing factor to this is that the threshold $\Sth$ is set
by assuming that the search at each binary orbital parameter is
independent,
while in fact there are significant correlations between adjacent points in
search parameter space. These correlations are difficult to safely account for
and so we make the conservative assumption that they are independent.
Thus $\Sth$
is an overestimate of the threshold for a one per cent false alarm
probability, in turn overestimating the upper limits and making a direct
comparison of the upper limits difficult.

This search also uses updated binary orbital parameter ranges, taking advantage
of a more
recent analysis of electromagnetic observations to produce a search better
targeted at Sco~X-1. Similarly, while the detector design is
fundamentally unchanged between O1 and O2, various detector improvements mean
that some instrumental lines have been removed or ameliorated, making this
search sensitive to signals that would have been obscured by instrumental noise
in searches using earlier datasets. The hidden Markov model is also
designed with particular emphasis on robustness to spin wandering.
Together, these three reasons mean that the search covers a slightly different
region of parameter space compared to previous Sco~X-1 searches.

\section{Conclusion}
In this paper, we search the LIGO O2 dataset for continuous gravitational waves from
the LMXB Sco~X-1, using a hidden Markov model combined with the \jstat{}.
We find no signal. The
search band extends from 60 Hz to 650 Hz.
The sky location $\alpha$, $\delta$ and orbital parameters $P$, $a_0$ and
$\phi_a$
used for the matched filter are electromagnetically constrained; values are given
in Table~\ref{tbl:scox1params}.
Monte-Carlo simulations of spin-wandering signals injected into the LIGO O2 data
imply frequentist 95\% upper limits of
$h_0^{95\%} = 3.47\times10^{-25}$,
$1.92\times10^{-25}$,
$1.42\times10^{-25}$
for
unknown, electromagnetically restricted
($\cos \iota
\approx 0.72$) and circular polarizations respectively.
The upper limits apply at 194.6 Hz, which is the most sensitive search
frequency.
For the electromagnetically-restricted case, the limit is $3.1$ times above, or
$1.2$ times below, the
torque-balance limit, when the torque-balance lever arm
is the stellar radius or the Alfv\'en radius respectively.
Monte-Carlo simulations are used to establish a detection threshold
corresponding to a false
alarm probability of $\alpha = 0.01$.

These results improve on the results from the previous HMM search,
described in Ref.~\citep{2017lscHMM}, by using data from LIGO's second
observing run, and by substituting the \jstat{} for the Bessel-weighted
\fstat{} to track the phase of the orbital Doppler shift.
As a result, the search in this paper is $\approx 2$ times more sensitive
compared to that in Ref.\citep{2017lscHMM}. The analysis remains
computationally efficient, requiring $\lesssim 3 \times 10^5$ GPU-hr for the
search itself and $\lesssim 10^6$ GPU-hr for simulations to
characterize the sensitivity and false alarm rate.

\section{Acknowledgements}

The authors gratefully acknowledge the support of the United States
National Science Foundation (NSF) for the construction and operation of the
LIGO Laboratory and Advanced LIGO as well as the Science and Technology Facilities Council (STFC) of the
United Kingdom, the Max-Planck-Society (MPS), and the State of
Niedersachsen/Germany for support of the construction of Advanced LIGO 
and construction and operation of the GEO600 detector. 
Additional support for Advanced LIGO was provided by the Australian Research Council.
The authors gratefully acknowledge the Italian Istituto Nazionale di Fisica Nucleare (INFN),  
the French Centre National de la Recherche Scientifique (CNRS) and
the Foundation for Fundamental Research on Matter supported by the Netherlands Organisation for Scientific Research, 
for the construction and operation of the Virgo detector
and the creation and support  of the EGO consortium. 
The authors also gratefully acknowledge research support from these agencies as well as by 
the Council of Scientific and Industrial Research of India, 
the Department of Science and Technology, India,
the Science \& Engineering Research Board (SERB), India,
the Ministry of Human Resource Development, India,
the Spanish  Agencia Estatal de Investigaci\'on,
the Vicepresid\`encia i Conselleria d'Innovaci\'o, Recerca i Turisme and the Conselleria d'Educaci\'o i Universitat del Govern de les Illes Balears,
the Conselleria d'Educaci\'o, Investigaci\'o, Cultura i Esport de la Generalitat Valenciana,
the National Science Centre of Poland,
the Swiss National Science Foundation (SNSF),
the Russian Foundation for Basic Research, 
the Russian Science Foundation,
the European Commission,
the European Regional Development Funds (ERDF),
the Royal Society, 
the Scottish Funding Council, 
the Scottish Universities Physics Alliance, 
the Hungarian Scientific Research Fund (OTKA),
the Lyon Institute of Origins (LIO),
the Paris \^{I}le-de-France Region, 
the National Research, Development and Innovation Office Hungary (NKFIH), 
the National Research Foundation of Korea,
Industry Canada and the Province of Ontario through the Ministry of Economic Development and Innovation, 
the Natural Science and Engineering Research Council Canada,
the Canadian Institute for Advanced Research,
the Brazilian Ministry of Science, Technology, Innovations, and Communications,
the International Center for Theoretical Physics South American Institute for Fundamental Research (ICTP-SAIFR), 
the Research Grants Council of Hong Kong,
the National Natural Science Foundation of China (NSFC),
the Leverhulme Trust, 
the Research Corporation, 
the Ministry of Science and Technology (MOST), Taiwan
and
the Kavli Foundation.
The authors gratefully acknowledge the support of the NSF, STFC, INFN, CNRS,
Swinburne University of Technology, the National Collaborative Research
Infrastructure Strategy of Australia, and the State of Niedersachsen/Germany
for provision of computational resources.

This work has been assigned LIGO document number LIGO-P1800208.

\bibliography{ms}

\end{document}